\documentclass[epj,nopacs]{svjour}
\usepackage{graphicx}
\usepackage{dcolumn}
\usepackage{bm}
\usepackage{multirow}
\usepackage{tabularx}
\usepackage{placeins}
\usepackage{subfigure}
\usepackage{subfig}
\usepackage{subfloat}
\usepackage{caption}
\usepackage{color}
\usepackage{ulem}
\usepackage[justification=justified]{caption}
\begin{document}
\title{Study of the $\boldsymbol{\mathsf{^{25}}}$Mg(d,p)$\boldsymbol{\mathsf{^{26}}}$Mg reaction to constrain the $\boldsymbol{\mathsf{^{25}}}$Al(p,$\boldsymbol{\gamma}$)$\boldsymbol{\mathsf{^{26}}}$Si resonant reaction rates in nova burning conditions}%

\author{C.~B.~Hamill\inst{1,}\thanks{e-mail: \texttt{conor.hamill@ed.ac.uk} (corresponding author)} \and 
P.~J.~Woods\inst{1} \and 
D.~Kahl\inst{1} \and 
R.~Longland\inst{2,3} \and
J.~P.~Greene\inst{4} \and
C.~Marshall\inst{2,3} \and
F.~Portillo\inst{2,3} \and
K.~Setoodehnia\inst{2,3,}\thanks{\emph{Present Address}: European X-ray Free Electron Laser GmbH, Holzkoppel 22869, Schenefeld, Germany} 
}

\institute{School of Physics \& Astronomy, The University of Edinburgh, James Clerk Maxwell Building, Edinburgh, EH9 3FD, UK \and	
Department of Physics, North Carolina State University, Raleigh, NC, 27695, USA \and
Triangle Universities Nuclear Laboratory, Duke University, Durham, NC, 27710, USA \and
Physics Division, Argonne National Laboratory, Argonne, IL 60439, USA
}

%
%
%
\date{Received: 23 October 2019/ Accepted: 26 December 2019}
%
\abstract{
The rate of the $^{25}$Al($p$,\,$\gamma$)$^{26}$Si reaction is one of the few key remaining nuclear uncertainties required for predicting the production of the cosmic $\gamma$-ray emitter $^{26}$Al in explosive burning in novae.
This reaction rate is dominated by three key resonances ($J^{\pi}=0^{+}$, $1^{+}$ and $3^{+}$) in $^{26}$Si.
Only the $3^{+}$ resonance strength has been directly constrained by experiment.
A high resolution measurement of the $^{25}$Mg($d$,\,$p$) reaction was used to determine spectroscopic factors for analog states in the mirror nucleus, $^{26}$Mg.
A first spectroscopic factor value is reported for the $0^{+}$ state at 6.256~MeV, and a strict upper limit is set on the value for the $1^{+}$ state at 5.691~MeV, that is incompatible with an earlier ($^{4}$He,\,$^{3}$He) study.
These results are used to estimate proton partial widths, and resonance strengths of analog states in $^{26}$Si contributing to the $^{25}$Al($p$,\,$\gamma$)$^{26}$Si reaction rate in nova burning conditions.
%
} 
\titlerunning{Study of the $^{25}$Mg($d$,\,$p$)$^{26}$Mg reaction to constrain the $^{25}$Al($p$,\,$\gamma$)$^{26}$Si resonant reaction rates\ldots}
\maketitle

\section{Introduction}
Astronomical observation of the characteristic 1809-keV $\gamma$ ray associated with the $\beta$-decay of the ground state of $^{26}$Al ($t_{1/2} = 7.17 \times 10^{5}$~yr) is one of the key pieces of evidence indicating stellar nucleosynthesis is an ongoing process in our galaxy.
Measurements of this spectral line by $\gamma$-ray telescopes have allowed the mass of $^{26}$Al in our galaxy to be progressively constrained to values of $2.8\pm0.8$ \cite{Diehl2006}, $2.7\pm0.7$ \cite{Wang2009} and $2.0\pm0.3$~$M_{\odot}$ \cite{Diehl2017}.
The $\gamma$-ray telescope \mbox{INTEGRAL} has localized the production of $^{26}$Al to known star-forming regions of our galaxy \cite{Martin2009}, where the main contributors are likely to be massive stars in their Wolf-Rayet and/or supernova phases \cite{Diehl2006,Smith2004,Prantzos1996}.
However, classical novae have received considerable attention as another potential source of this radioisotope and have been estimated to contribute significantly to the amount in our galaxy, with theoretical values of up to $0.4$ \cite{Jose1997} and $0.8$~$M_{\odot}$ \cite{Bennett2013} previously calculated.
Extinct $^{26}$Al is observed in the form of high abundances of the isotope $^{26}$Mg \textcolor{black}{($\beta$-decay daughter of $^{26}$Al)} in presolar grains originating from a single nova event \cite{Amari2001,Amari2002}.
An outstanding issue in nova models is that the calculated ejecta require mixing with solar-like material prior to grain condensation \cite{Jose2004}.
The resolution of this problem could lie within the models themselves, including the nuclear physics input data, or with the interpretations of the observations.

Classical novae involve a thermonuclear runaway and the ejection of material whenever a white dwarf in a binary stellar system has accreted sufficient material from its companion star \cite{Starrfield2016} and have been predicted to occur at a galactic rate of $50^{+31}_{-23}$ per year \cite{Shafter2017}.
$^{26}$Al is produced by a series of proton capture reactions and $\beta$-decays during explosions that reach temperatures in the range of $0.1$--$0.4$~GK \cite{Iliadis2002}.
At high temperatures, the $^{25}$Al($p$,\,$\gamma$)$^{26}$Si reaction rate can become faster than $^{25}$Al $\beta$-decay.
In this scenario $^{26}$Si subsequently $\beta$-decays to the short lived $0^{+}$ isomeric state of $^{26}$Al, leading to the bypassing of the production of the ground state. The $^{25}$Al($p$,\,$\gamma$)$^{26}$Si reaction rate at nova burning temperatures is expected to be dominated by three resonances in $^{26}$Si corresponding to excitation energies of 5.676 (spin/parity $J^{\pi}=1^{+}$), 5.890 ($0^{+}$) and 5.929 ($3^{+}$) MeV. 

However, direct measurements of the individual resonance strength contributions to the $^{25}$Al($p$,\,$\gamma$)$^{26}$Si reaction rate are not feasible with presently available $^{25}$Al radioactive beam intensities.
Therefore indirect approaches are required to constrain these rates.
For the $3^{+}$ state in $^{26}$Si, corresponding to an s-wave resonance, a measurement of its proton decay in the $^{25}$Al($d$,\,$n$)$^{26}$Si reaction, was used to estimate the proton partial width, $\Gamma_p$, of the state \cite{Peplowski2009}.
A subsequent $\beta$-decay study of $^{26}$P \cite{Bennett2013} measured the $\gamma$-decay branching ratio of the $3^{+}$ state, enabling a value for the resonance strength to be derived, $\omega\gamma= 23\pm6({\rm stat.})$~meV.
No experimental information is available to similarly constrain the strength contributions for the $1^{+}$ and $0^{+}$ states in $^{26}$Si.
Here we consider an alternate approach exploring single particle strengths of analog states at 5.691 ($1^{+}$), 6.125 ($3^{+}$) and 6.256 ($0^{+}$) MeV in $^{26}$Mg produced by the $^{25}$Mg($d$,\,$p$) reaction.
In earlier studies of this reaction (see refs.~\cite{Burlein1984,Arciszewski1984a}) spectroscopic factor values were only reported for the $3^{+}$ state.
For the analog $1^{+}$ and $0^{+}$ states in $^{26}$Si the proton partial width is predicted by shell-model calculations to be \textcolor{black}{comparable to or} weaker than the $\gamma$-width \cite{Richter2011} and therefore will strongly affect the resonance strength.
It is therefore important that spectroscopic strengths for these states be constrained experimentally.
It has been noted for example that in a study of the $^{25}$Mg($^{4}$He,\,$^{3}$He)$^{26}$Mg reaction \cite{Yasue1990a} a spectroscopic factor value was reported for the $1^{+}$ state a factor $\sim\!50$ larger than shell model predictions \cite{Parikh2013}.
\section{Experimental Setup}
\begin{figure}
\includegraphics[scale=0.48]{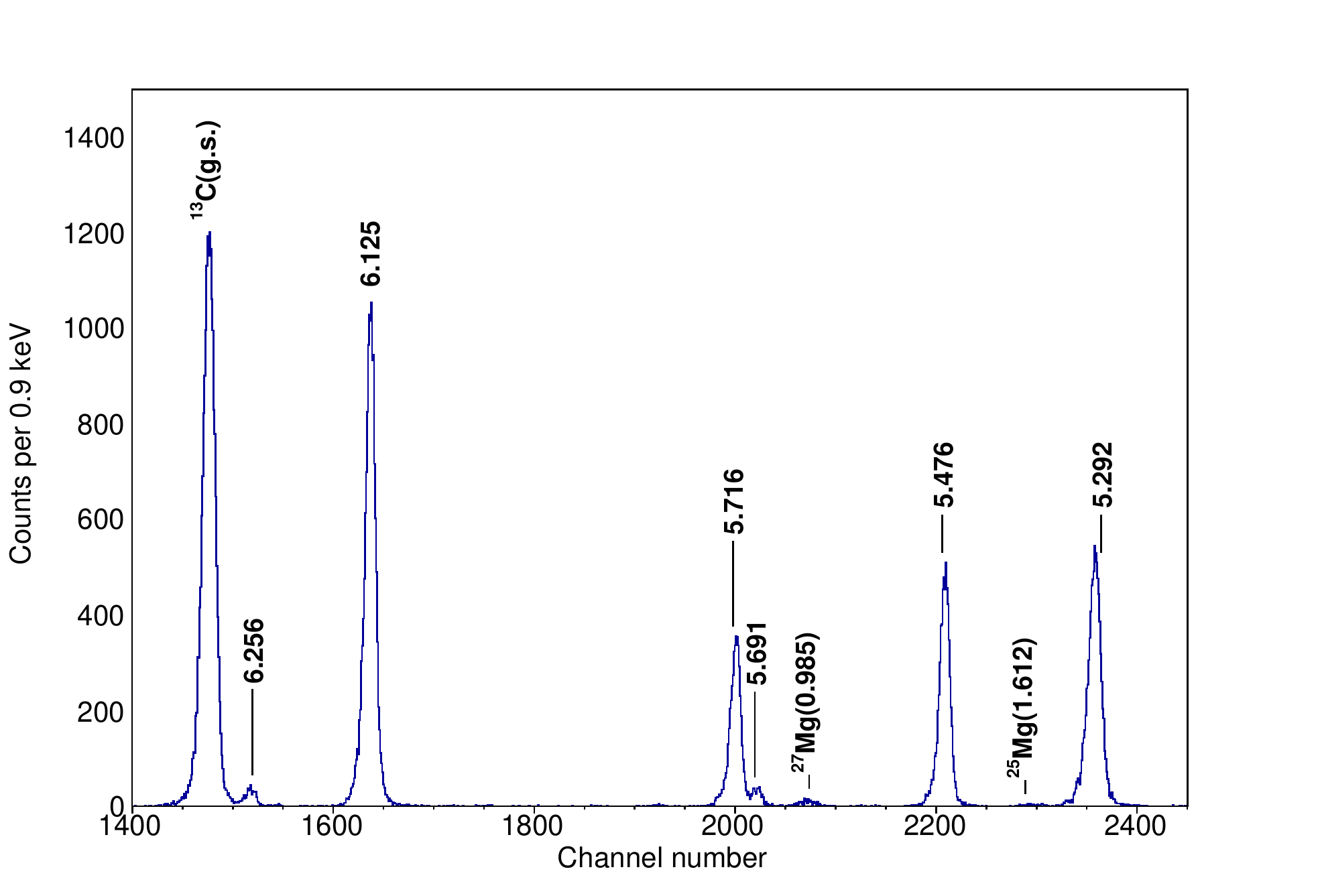} 
\captionsetup{format=plain}
\caption{Energy spectrum from the $^{25}$Mg($d$,\,$p$)$^{26}$Mg reaction at $\mathrm{\theta_{lab}=30^{\circ}}$. Excited $\mathrm{^{26}Mg}$ states are labelled with their excitation energies in MeV, taken from ref.~\cite{Basunia2016}. Contaminant peaks are labelled with their corresponding final excited states.}
\label{fig:spectrum_zoomed}
\end{figure}

The present experiment was performed at the Triangle Universities Nuclear Laboratory (TUNL).
A deuteron beam was accelerated to an energy of 8~MeV
by the 10-MV FN tandem Van de Graaff accelerator.
The beam was momentum analyzed by the high-resolution beam line at TUNL using two $90^{\circ}$ magnets.
Beam currents on target varied between $\sim\!250$--$800$~enA and were measured using a suppressed beam stop positioned at zero degrees.
The $^{25}$Mg($d$,\,$p$) reaction was measured using $^{25}$Mg targets enriched to a nominal isotopic composition of $99.2\pm0.1\%$.
The two targets used had thicknesses of 90 and 112~$\mu$g/cm$^{2}$, measured using alpha particle energy loss (estimated uncertainty $\pm10\%$), and were backed by a thin gold flash. 
The TUNL high resolution Enge split-pole magnetic spectrograph accepted protons from the reactions, with an opening angle of 1.0~msr, which were then focused on to the spectrograph's focal plane.
The positions of the momentum-dispersed particles on the focal plane were measured using two position sensitive avalanche counters. 
A $\Delta E/E$ detector combination, consisting of a gas proportionality counter to measure energy deposited, and a residual energy scintillator to measure total energy, allowed discrimination between different species of light ions.
The detector system is described in greater detail in ref.~\cite{Marshall2019}.
Measurements of the reaction products were taken at multiple angles between $13$--$55^{\circ}$. \sloppy 
The beam energy was chosen to separate the protons produced strongly via the $^{12}$C($d$,\,$p$)$^{13}$C reaction from those corresponding to the 6.256~MeV $0^{+}$ peak at more forward angles.
Excited states corresponding to the key states at 5.691, 6.125 and 6.256~MeV in $^{26}$Mg (see Fig.~\ref{fig:spectrum_zoomed}) were identified by a polynomial fit to well-known, strongly produced, states in $^{26}$Mg, with all observed peaks matching a known level of $^{26}$Mg within 5~keV \cite{Basunia2016}.
All peaks shown in the excitation energy spectrum of Fig.~\ref{fig:spectrum_zoomed} were identified either as corresponding to known states in $^{26}$Mg or weak isotopic target contamination peaks corresponding to known states in $^{25}$Mg, $^{27}$Mg, or $^{13}$C. 
The energy resolution (FWHM), was around 14--16~keV across all angles.
No evidence for peak broadening was observed during the runs, indicating there was no significant target degradation over time. This was also checked by monitoring the yields of strongly produced peaks against the integrated beam current.

\begin{table*}
\centering
\caption{Optical model potential parameters used in DWBA analysis of the $^{25}$Mg($d$,\,$p$)$^{26}$Mg reaction. The first two potential sets refer to the entrance and exit scattering channels, the third refers to the core-core interaction and the fourth describes the binding potential of the residual nucleus. Parameters have meanings as defined in ref.~\cite{Koning2003}, with energies in MeV and distances in fm.}
\begin{tabular}{cccccccccccccc}
\hline\noalign{\smallskip}
 Potential &$V_{v}$ &$r_{v}$ &$a_{v}$ &$W_{v}$ &$r_{wv}$ & $a_{wv}$ &$W_{D}$ &$r_{D}$ &$a_{D}$ &$V_{so}$ &$r_{so}$ & $a_{so}$ & $r_{c}$ \\ \noalign{\smallskip}\hline\noalign{\smallskip}

\rule{0pt}{2.5ex}$^{25}\mathrm{Mg} + d$ \cite{Han2006} &83.9&1.17&0.81 &0.0 &1.56 &0.83 & 18.6& 1.33& 0.60 & 3.70 & 1.23 & 0.81 & 1.70 \\

$^{26}\mathrm{Mg} + p$ \cite{Koning2003} &53.7&1.17&0.67 &0.64 & 1.17 & 0.67 & 8.02& 1.34 & 0.53 & 5.69 & 0.97 & 0.59 & 1.33 \\

$^{25}\mathrm{Mg} + p$ \cite{Koning2003} & 53.7 & 1.17 & 0.67 & 0.64 & 1.17 & 0.67 & 8.02 & 1.34 & 0.53 & 5.69 & 0.97 & 0.59 & 1.33 \\

$^{25}\mathrm{Mg} + n$ \cite{Soukhovitski2005} &52.1 & 1.16 & 0.64 & -- & -- & -- &  -- & -- & -- & 5.50 & 0.96 &0.59 & 1.26   \\
\noalign{\smallskip}\hline
\end{tabular}
\label{OMPlabel}
\end{table*}

\section{Results and Analysis}

\begin{figure}
	\centering
	\begin{tabular}[c]{c}
		\begin{subfigure}{}
				\includegraphics[width=67mm]{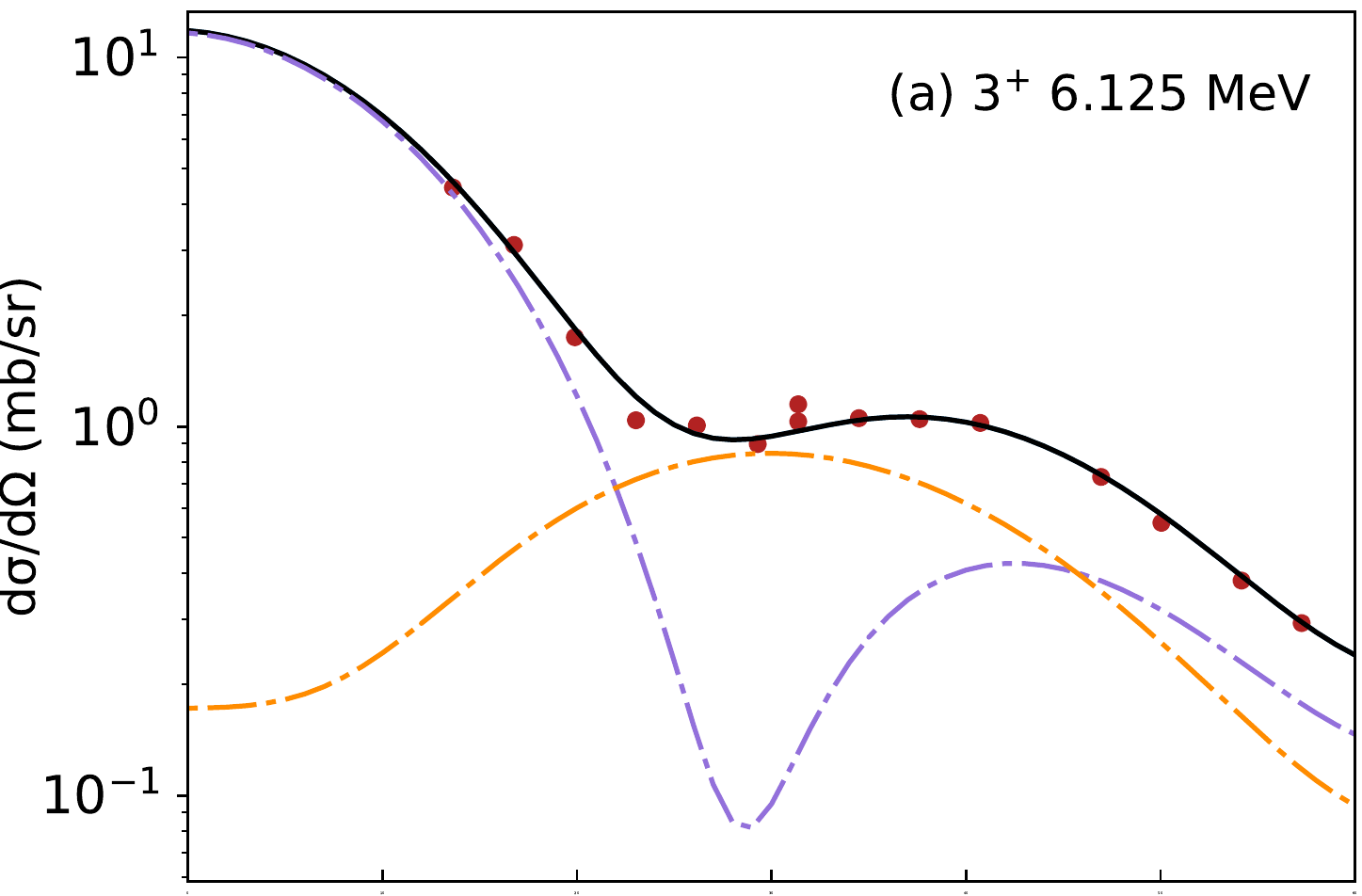}
\label{fig:test label}
		\end{subfigure}\\
		\begin{subfigure}{}
				\includegraphics[width=67mm]{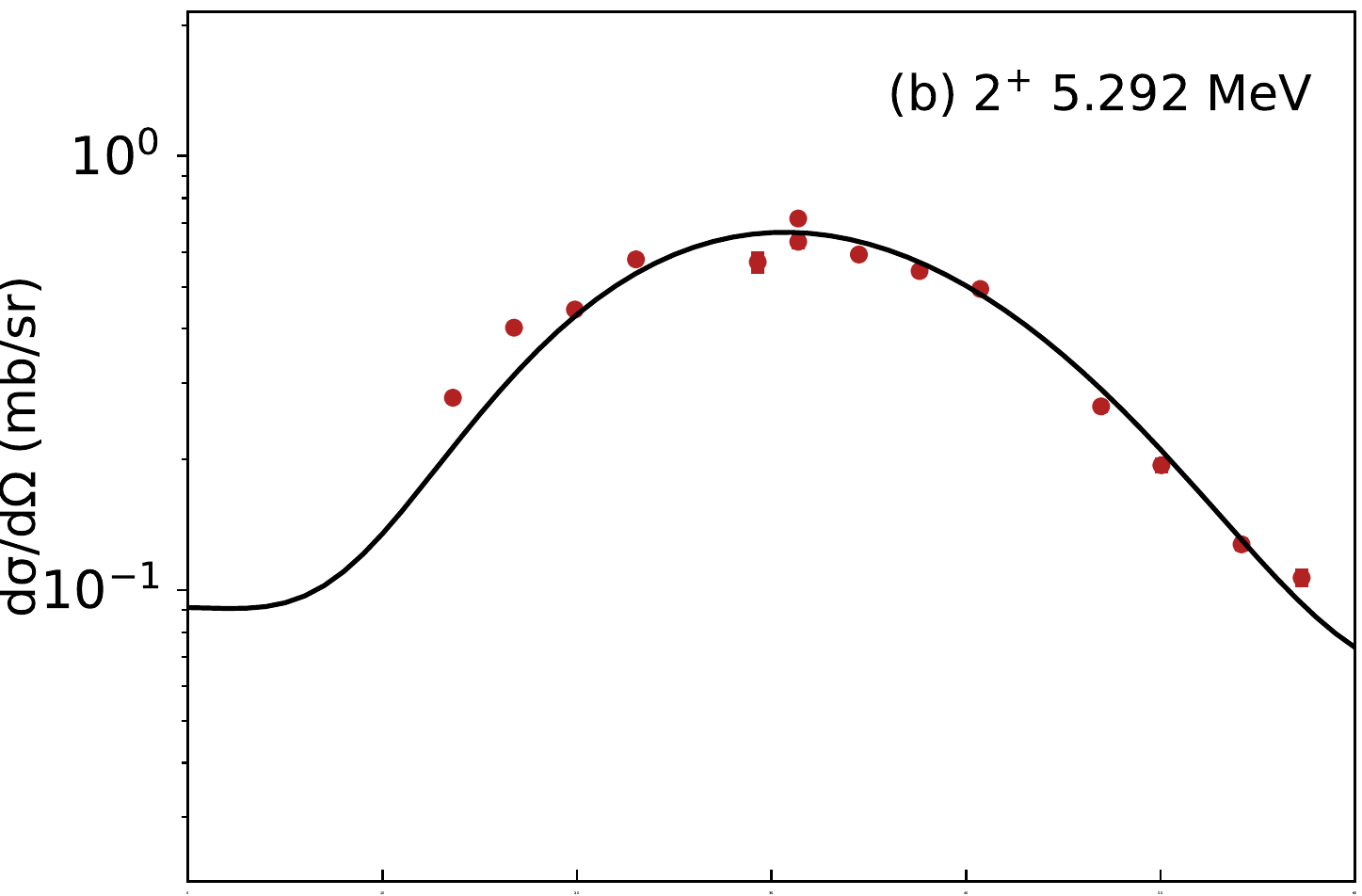}
\label{fig:test label}
		\end{subfigure}\\
				\begin{subfigure}{}
				\includegraphics[width=67mm]{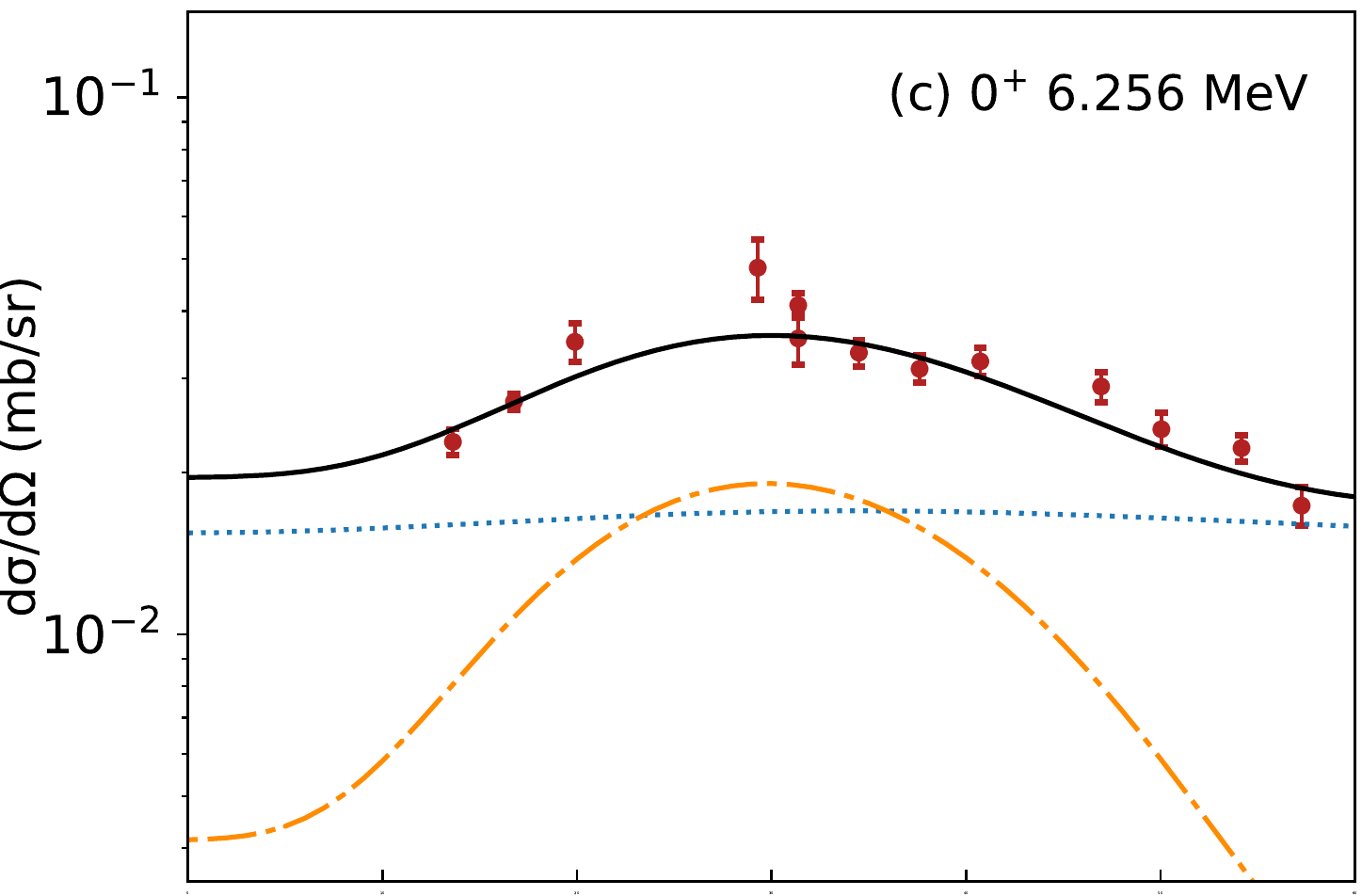}
\label{fig:test label}
		\end{subfigure}\\
		\begin{subfigure}{}
				\hspace{0.2mm}
				\includegraphics[width=68.3mm]{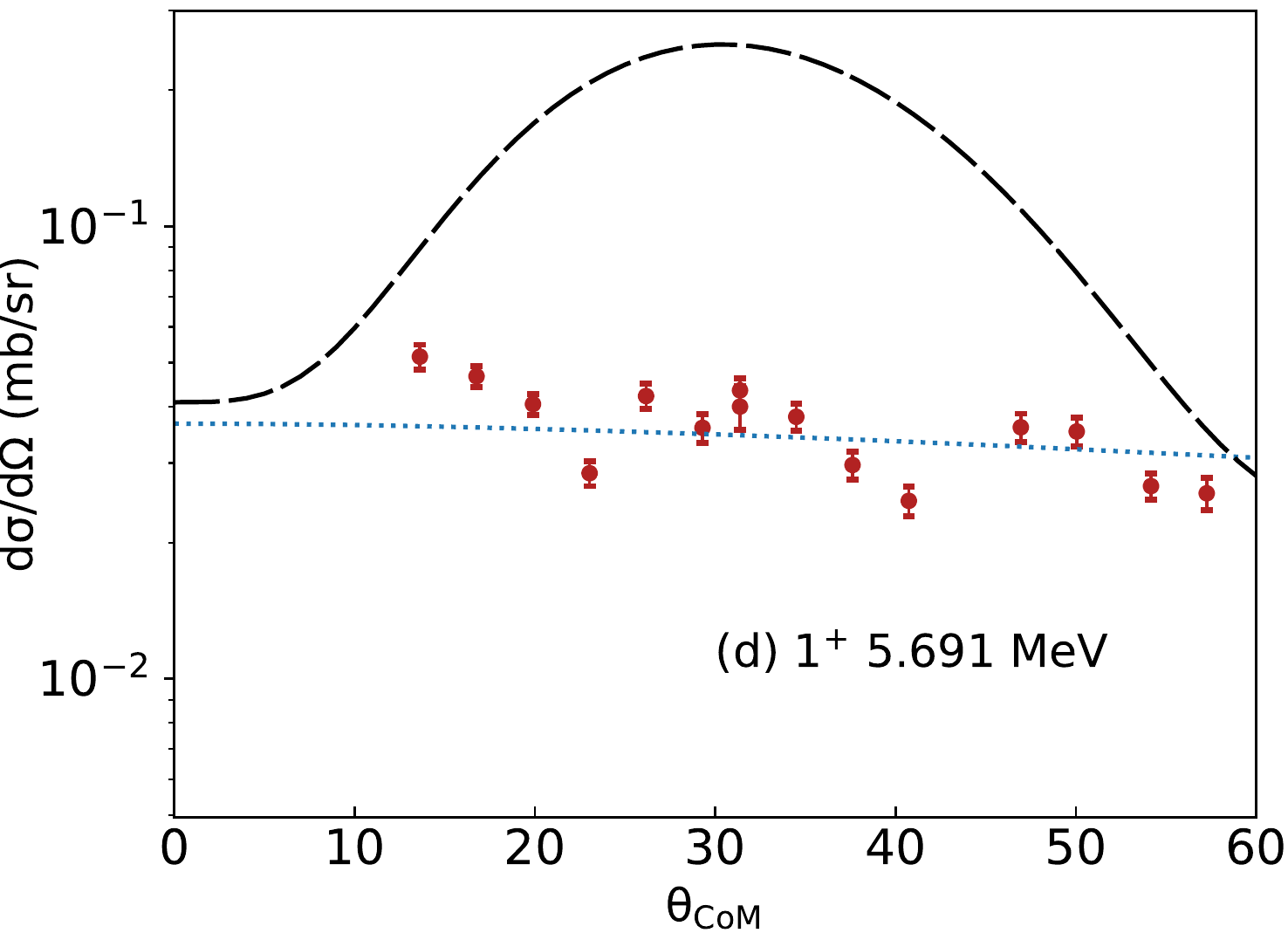}
\label{fig:test label}
		\end{subfigure}		
	\end{tabular}
\caption{Differential cross-section measurements for the $^{25}$Mg($d$,\,$p$)$^{26}$Mg reaction. In a) for the 6.125~MeV state, the solid black line represents a fit using \textsc{fresco} calculations combining $\ell=0$ (purple) and $\ell=2$ (orange) components. For comparison the 5.292~MeV state is shown in b) fitted with just an $\ell=2$ component (black line). In c) the total fit for the 6.256~MeV state (black line) requires both a direct $\ell=2$ component (orange line) and a compound nuclear component (blue) to reproduce the angular distribution. For the 5.691~MeV state shown in d), there is no evidence for a direct $\ell=2$ component: the dashed black line represents the distribution expected if the $C^{2}S(\ell=2)$ value taken from the ($^{4}$He,\,$^{3}$He) study \cite{Yasue1990a} is adopted in \textsc{fresco} calculations, while the blue line shows the angular distribution (scaled to fit the data) predicted assuming a dominant compound nuclear mechanism.}
\label{ang_dist_label}	
\end{figure}
Figure~\ref{ang_dist_label} shows the experimentally-measured angular distributions of states at 5.292 ($2^{+}$), 5.691, 6.125 and 6.256~MeV in $^{26}$Mg.
The errors on the individual data points are statistical uncertainties.
These data are compared with angular distributions calculated using the distorted wave Born approximation (DWBA) with the \textsc{fresco} program \cite{Thompson1988}.
The $p + n$ interaction was described by a Gaussian potential (see refs.~\cite{Austern1987,Avrigeanu2010}). The parameters chosen were a depth of 72.2~MeV and a root mean square value of 1.48~fm, determined in ref.~\cite{Kamimura1986} by the fitting of the potential to reproduce the deuteron binding energy.
For the other interactions, real and imaginary volume, imaginary surface potential and a real spin-orbit potential, a Woods-Saxon shape was used. The depth of the central potential was varied to produce the correct binding energy of the excited states of $^{26}$Mg.
The potential parameter sets used are listed in Table~\ref{OMPlabel}.
The following expression was used to calculate spectroscopic factors, $C^{2}S$:
\begin{equation}
\frac{d\sigma}{d\Omega}_{\rm exp}=C^{2}S\frac{d\sigma}{d\Omega}_{\rm th}.
\end{equation}
Table~\ref{specfactorslabel} shows the present experimental $C^{2}S$ values compared with those obtained in $^{25}$Mg($d$,\,$p$) reaction studies by Burlein \textit{et al.} \cite{Burlein1984}, Arciszewski \textit{et al.}  \cite{Arciszewski1984a}, the $^{25}$Mg($^4$He,\,$^3$He) study of Yasue \textit{et al.} \cite{Yasue1990a}, and a shell model calculation \cite{Richter2011}.  Here we estimate uncertainties in the derived  $C^{2}S$ values based on a combination of the overall goodness of fit of the angular distribution, the experimental cross-section normalisation uncertainty (10\%), and the uncertainty in the choice of optical model parameters (estimated here to be 20\% from a consideration of different available theoretical parameter sets, for example refs.~\cite{An2006,Li2008}).  

We consider first the strongly produced $3^{+}$ state at 6.125~MeV. 
The angular distribution (see Fig.~\ref{ang_dist_label}a) clearly requires both orbital angular momentum $\ell=0$ and $\ell=2$ components for a good fit of the distribution.
This can be contrasted with the state at 5.292~MeV which has been associated with a relatively pure $\ell=2$ component in earlier work \cite{Burlein1984,Arciszewski1984a}, consistent with what we also observe here (see for comparison Fig.~\ref{ang_dist_label}b).
The $C^{2}S(\ell=0)$ value for the 6.125~MeV state we obtain is in excellent agreement with the values reported by both Burlein \textit{et al.} \cite{Burlein1984} and Arciszewski \textit{et al.} \cite{Arciszewski1984a}.
The $C^{2}S(\ell=2)$ value reported here is broadly consistent with, but larger than, the value reported by Burlein \textit{et al.} \cite{Burlein1984}, but a factor of \textcolor{black}{$\sim\!2$} smaller than reported in ref.~\cite{Arciszewski1984a}.
The $C^{2}S$ values for both the $\ell=0$ and $\ell=2$ components from the present experiment agree well with the shell-model predictions and the ($^{4}$He,\,$^{3}$He) study of Yasue \textit{et al.} \cite{Yasue1990a}.

The $0^{+}$ state at 6.256~MeV is populated by $\ell=2$ transfer.
The \textsc{fresco} calculations reproduce the peak observed at $\sim\!\!30^{\circ}$ (see Fig.~\ref{ang_dist_label}c) but the peak is less pronounced than calculations predict. At these relatively low beam energies the compound nuclear reaction mechanism can potentially contribute significantly to the total ($d$,\,$p$) reaction cross section, particularly for states less strongly produced by the direct transfer mechanism.
In Fig.~\ref{ang_dist_label}c we show an angular distribution for this state calculated using \textsc{talys} \cite{Koning2007} based on the Hauser-Feshbach approach to the compound nucleus mechanism.
It is essentially flat.
A good fit to the data can be obtained adding the direct and compound mechanisms together and allowing the magnitudes of the two components of the cross section to be variable (this is equivalent to assuming energy-averaged fluctuations are approximately zero for the compound nuclear component \cite{Hodgson1971}---this analysis approach is used in refs.~\cite{Gallmann1966,Schmick1974,Meurders1975} in studies of the $^{24}$Mg($d$,\,$p$) reaction, for example).
Using this method, a first value for $C^{2}S(\ell=2)$ can be obtained for the ($d$,\,$p$) reaction.
This value agrees very well both with the value from the $^{25}$Mg($^{4}$He,\,$^{3}$He) study of Yasue \textit{et al.} \cite{Yasue1990a}, and the shell-model calculation \cite{Richter2011}, suggesting a relatively weak single particle component compared to the 3\textsuperscript{+} state. 

We now consider the $1^{+}$ state at 5.691~MeV.  
In Fig.~\ref{ang_dist_label}d a calculated cross section is shown assuming a $C^{2}S(\ell=2)$ value of 0.20 taken from the $^{25}$Mg($^{4}$He,\,$^{3}$He) study of Yasue \textit{et al.} \cite{Yasue1990a}.
The experimental angular distribution is completely incompatible with the direct transfer reaction calculation.
However, the shape of the angular distribution is compatible with a single dominant compound nuclear mechanism for populating this state (see Fig.~\ref{ang_dist_label}d). 
An upper limit (at the 1$\sigma$ confidence level) obtained on the $C^{2}S$ value for the direct component is small, but consistent with shell model predictions \cite{Richter2011} (see Table~\ref{specfactorslabel}). 
Yasue \textit{et al.} suggested in their own work that large multistep reaction processes may cause higher yields for the ($^{4}$He,\,$^{3}$He) reaction to $1^{+}$ states \cite{Yasue1990a}, and there was difficulty resolving this state from a neighbouring $4^{+}$ state at 5.72~MeV in $^{26}$Mg.  
Burlein \textit{et al.}, also had difficulty resolving these states in their ($d$,\,$p$) study and only quoted a spectroscopic factor value for the doublet (see Table~II in ref. \cite{Burlein1984}). \textcolor{black}{Higher-order and multistep mechanisms are not treated in the present reaction analysis.}

\begin{figure}[b]
\includegraphics[scale=0.46]{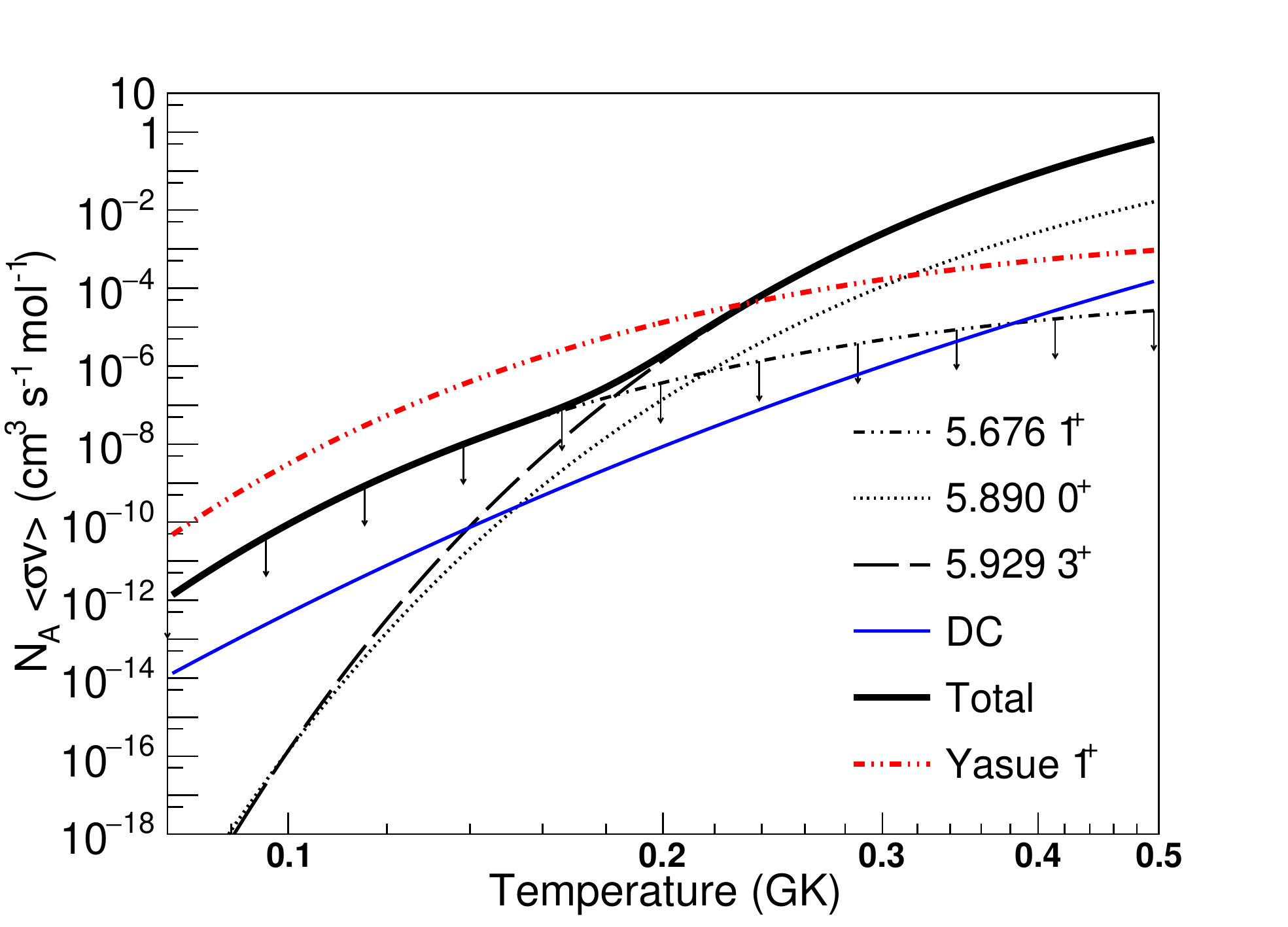}
\caption{Calculated rate of the $^{25}$Al($p$,\,$\gamma$)$^{26}$Si reaction using the parameters given in Table~\ref{tab:widthslabel} for nova burning. The arrows pointing downwards for the contribution of the 1\textsuperscript{+} resonance correspond to an upper limit. The dashed red line shows the corresponding rate for the 1\textsuperscript{+} resonance rate if the spectroscopic value taken from the Yasue study of the ($^{4}$He,\,$^{3}$He) reaction is adopted \cite{Yasue1990a}.}
\label{reactionratelabel}
\end{figure}

\begin{table*}	
\centering
\caption{Neutron spectroscopic factors of states of interest in \textsuperscript{26}Mg measured in this experiment, compared to values obtained in previous studies. Also shown are shell model calculations of proton spectroscopic factors for corresponding analog states in \textsuperscript{26}Si, relevant for the $^{25}$Al($p$,\,$\gamma$)$^{26}$Si reaction in novae.}
\begin{tabular}{cccccccc}
\hline\noalign{\smallskip}
 &	& 	 &\multicolumn{4}{c}{$C^2S_\mathrm{{exp}}$}&$C^2S\mathrm{_{th}}$\\ \noalign{\smallskip}\cline{4-7}\noalign{\smallskip}
 \rule{0pt}{9pt}
$E\mathrm{_x}$~(MeV) \cite{Basunia2016}&$J^{\pi}$&$\ell$& $(d,p)$ \cite{Burlein1984} &$(d,p)$ \cite{Arciszewski1984a}
& $(^4\mathrm{He},^3\mathrm{He})$ \cite{Yasue1990a}  &Current Work & Shell Model \cite{Richter2011}\\ 
\noalign{\smallskip}\hline\noalign{\smallskip}
 
  5.69108(19)&$1^{+}$&2&  &  & 0.20(4) & $\mathrm{<}5.7\times10^{-3}$~$^{a}$& $3.5\times10^{-3}$ \\

 6.12547(5)&$3^{+}$&0, 2 & 0.121, 0.206~$^{b}$& 0.106(13), 0.60(14)&    0.14(3), 0.30(6)&  0.11(2), 0.27(6) & 0.14, 0.33\\

 6.25547(5)&$0^{+}$&2&		&	& 0.054(11)& 0.042(10) &0.039\\
\noalign{\smallskip}\hline
\end{tabular}
\label{specfactorslabel}
\begin{flushleft}
$^{a}$   Upper limit at 1$\sigma$ confidence level. \\
$^{b}$   No uncertainties were provided in this reference.
\end{flushleft}
\end{table*}

\begin{table*}
\centering
\caption{Resonance parameters used to calculate the $^{25}$Al($p$,\,$\gamma$)$^{26}$Si reaction rate shown in Fig.~3 (see text for more details). Resonance energies have been calculated using a proton separation energy for \textsuperscript{26}Si, $S_p\mathrm{=5.51401(11)~MeV}$ \cite{Huang2017}.}

\newcommand\Tstrut{\rule{0pt}{2.6ex}}         
\newcommand\Bstrut{\rule[-0.9ex]{0pt}{0pt}}   

\begin{tabular}{cccccc}
\hline\noalign{\smallskip}
$E\mathrm{_x}$~(MeV) \cite{Basunia2016} & $E\mathrm{_r}$~(MeV)\cite{Basunia2016,Huang2017} & $J^{\pi}$ & $\Gamma_p$~(eV) & $\Gamma_\gamma$~(eV) & $\omega\gamma$~(eV) \\ 
\noalign{\smallskip}\hline\noalign{\smallskip} 

5.6762(3) & 0.1622(3) & 1\textsuperscript{+} &     $\mathrm{<}1.0\times10^{-8}$
& 0.12~$^{a}$  & $\mathrm{<}2.6	\times10^{-9}$ \\

5.8901(3) & 0.3761(3) & 0\textsuperscript{+} &  $4.2\times10^{-3}$
& $8.8\times10^{-3}$~$^{a}$
& $2.4\times10^{-4}$ \\

5.9294(8)& 0.4154(8) & 3\textsuperscript{+} & 2.9~$^{b}$& 0.040~$^{c}$	
 & $2.3\times10^{-2}$ \\
\noalign{\smallskip}\hline
\end{tabular}
\label{tab:widthslabel}
\begin{flushleft} 
~~~~~~~~~~~~~~~~~~~~~~~~$^{a}$ \cite{Richter2011}. \\
~~~~~~~~~~~~~~~~~~~~~~~~$^{b}$ \cite{Peplowski2009}. \\
~~~~~~~~~~~~~~~~~~~~~~~~$^{c}$ \cite{Bennett2013}.
\end{flushleft}
\end{table*}

\section{$\boldsymbol{\mathsf{^{25}}}$A\MakeLowercase{l}(p,\,$\boldsymbol{\gamma}$)$\boldsymbol{\mathsf{^{26}}}$S\MakeLowercase{i} Reaction Rate}
As noted above, only the strength of the $3^{+}$ resonance in the $^{25}$Al($p$,\,$\gamma$)$^{26}$Si reaction rate is currently directly constrained by experiment \cite{Bennett2013,Peplowski2009}.
In their $^{25}$Al($d$,\,$n$) study, Peplowski \textit{et al.} assign a `large spectroscopic factor' to the $\ell=0$ component of the $3^{+}$ state \cite{Peplowski2009} and `based on [their] experimental cross-section' derive a proton partial width of $2.9(10)$~eV (there is some uncertainty due to a possible $\ell=2$ contribution to this state and from the unresolved $0^{+}$ resonance in the data). 
The present data set, and the earlier single neutron transfer data sets \cite{Burlein1984,Arciszewski1984a,Yasue1990a} give consistent values for $C^{2}S(\ell=0)$ for the analog $3^{+}$ state in the mirror nucleus $^{26}$Mg.
Using our present $C^{2}S(\ell=0)$ value for $^{26}$Mg, and scaling from the calculations of Richter \textit{et al.} \cite{Richter2011}, and assuming isospin symmetry, we would estimate a proton partial width of $\sim\!2.6$~eV for the $3^{+}$ resonance in the $^{25}$Al($p$,\,$\gamma$)$^{26}$Si reaction , consistent with the value derived by Peplowski \textit{et al.} \cite{Peplowski2009}. \textcolor{black}{In Table 3, we have adopted the value of the proton partial width of the 3$^{+}$ state deduced by Peplowski \textit{et al.} for the 3$^{+}$ resonance in $^{26}$Si.}
Taking the same calculational approach \textcolor{black}{as for the $3^{+}$ state}, we can use our new measurements on the $0^{+}$ and $1^{+}$ states to estimate their partial proton widths in the mirror nucleus $^{26}$Si. These values are shown in Table~\ref{tab:widthslabel}
 along with shell model calculations of their gamma partial widths taken from Richter \textit{et al.} \cite{Richter2011}.
The derived resonance strength value (for the $0^{+}$ state) and upper limit (for the $1^{+}$ state) are used for the $^{25}$Al($p$,\,$\gamma$)$^{26}$Si reaction rate calculation for nova burning temperatures shown in Fig.~3
\textcolor{black}{(previous $\Gamma_p$ estimates, e.g. \cite{Peplowski2009,Parpottas2004}, were based on shell-model calculations \cite{Iliadis1996})}.
For the excitation energy of the $0^{+}$ state, we have used the value of 5.890~MeV adopted in the most recent data compilation \cite{Basunia2016}, based on several recent $\gamma$-decay measurements \cite{DeSereville2010,Komatsubara2014,Doherty2015a}, to derive the resonance energy and estimate the proton partial width (an earlier ($^{3}$He,\,$n$) neutron time-of-flight measurement had assigned the $0^{+}$ state an excitation energy of 5.946~MeV \cite{Parpottas2004}).

The $3^{+}$ resonance reaction rate calculation uses the resonance strength value derived directly from information on the state in $^{26}$Si by Bennett \textit{et al.} \cite{Bennett2013}.
The direct capture \textcolor{black}{(DC)} contribution to the reaction rate of $^{25}$Al($p$,\,$\gamma$)$^{26}$Si was calculated using the approach outlined in ref. \cite{Matic2010} 
\textcolor{black}{(a total \textit{S}-factor of 28~keV-b was used; the USDA interaction was used to calculate $C^2S$ values).}
Considering the lower temperature regime below $T\sim0.2$~GK, it is the upper limit on the strength of the $1^{+}$ resonance that constrains the reaction rate. 
The $1^{+}$ reaction rate contribution implied by the much higher $C^{2}S(\ell=2)$ value from the ($^{4}$He,\,$^{3}$He) study of Yasue \textit{et al.} \cite{Yasue1990a} is also shown for comparison. 
Parikh and Jos\'e have calculated that even using the high strength value implied by Yasue \textit{et al.} the rate of destruction of $^{25}$Al under nova burning conditions up to $T\sim 0.2$~GK will be dominated by its $\beta$-decay rate \cite{Parikh2013}. 
Our significantly reduced upper limit on the $1^{+}$ strength reported here would further strengthen this conclusion.
The new value for the resonance strength derived for the $0^{+}$ state shows that this contributes $\sim\!10\%$ to the total $^{25}$Al($p$,\,$\gamma$)$^{26}$Si reaction rate at temperatures above 0.2~GK which is dominated by $\ell=0$ resonance capture on the $3^{+}$ state.   

\section{Summary}
In this paper we have presented the results of a $^{25}$Mg($d$,\,$p$)$^{26}$Mg experimental reaction study performed at TUNL using the Enge split-pole spectrometer. 
Our aim has been to study analog states of the three key resonances determining the $^{25}$Al($p$,\,$\gamma$)$^{26}$Si reaction rate in nova burning conditions. 
While the $3^{+}$ resonance strength contribution has experimental constraints, the strengths of the $0^{+}$ and $1^{+}$ resonances have not been similarly constrained. 
From our study we have been able to measure a first value of the spectroscopic factor for the ($d$,\,$p$) reaction to the $0^{+}$ state in $^{26}$Mg. 
From this value we were able to make an estimate of the proton partial width of the analog state in $^{26}$Si, assuming isospin symmetry. 
The value agrees well with shell model predictions. 
We conclude that in the nova burning region above a temperature $\sim\!0.2$~GK this produces a $\sim\!10\%$ contribution to the total reaction rate, which will be dominated by the contribution from the $3^{+}$ state. 
We have set a strict upper limit on the spectroscopic factor for the $1^{+}$ state in $^{26}$Mg, which is much smaller than the value previously deduced from a ($^{4}$He,\,$^{3}$He) reaction study \cite{Yasue1990a}. 
This discrepancy may be due to problems with additional multistep reaction contributions to the cross-section specifically for $1^{+}$ states as suggested in ref. \cite{Yasue1990a}, and/or due to the presence of a more strongly produced unresolved state in that study.  

The present stricter constraint on the upper limit for the $1^{+}$ resonance strength would indicate that the $^{25}$Al($p$,\,$\gamma$)$^{26}$Si reaction rate below $0.2$~GK in novae is likely to be dominated by $\beta$-decay \cite{Parikh2013}.  
Having reduced large uncertainties in the reaction rate contributions from the $0^{+}$ and $1^{+}$ resonances we therefore conclude that further efforts to constrain the $^{25}$Al($p$,\,$\gamma$)$^{26}$Si reaction rate in novae should concentrate on uncertainties in the contribution of the $3^{+}$ resonance.
\FloatBarrier
\section*{Acknowledgements}
The authors would like to thank the TUNL technical staff for their contributions. C.B.H, P.J.W. and D.K. would like to thank the UK STFC for support. This material is based upon work supported by the U.S. Department of Energy, Office of Science, Office of Nuclear Physics, under Award Number DE-SC0017799 and under Contract Nos. DE-FG02-97ER41041 and DE-AC02-06CH11357. C.B.H. would like to thank Antonio Moro for assistance with running \textsc{fresco} and Arjan Koning for clarification with input files for \textsc{talys}.

\bibliographystyle{prsty}
\bibliography{library}

\begin{thebibliography}{10}

\bibitem{Diehl2006}
R. Diehl {\it et~al.}, Nature {\bf 439},  45  (2006).

\bibitem{Wang2009}
W. Wang {\it et~al.}, Astron. Astrophys. {\bf 496},  713  (2009).

\bibitem{Diehl2017}
R. Diehl, JPS Conf. Proc. {\bf 14},  010302  (2017).

\bibitem{Martin2009}
P. Martin, J. Kn{\"{o}}dlseder, R. Diehl, and G. Meynet, Astron. Astrophys.
  {\bf 506},  703  (2009).

\bibitem{Smith2004}
D.~M. Smith, New Astron. Rev. {\bf 48},  87  (2004).

\bibitem{Prantzos1996}
N. Prantzos and R. Diehl, Phys. Rep. {\bf 267},  1  (1996).

\bibitem{Jose1997}
J. Jos{\'{e}}, Astrophys. J. {\bf 479},  55  (1997).

\bibitem{Bennett2013}
M.~B. Bennett {\it et~al.}, Phys. Rev. Lett. {\bf 111},  1  (2013).

\bibitem{Amari2001}
S. Amari, E. Zinner, J. Jose, and M. Hernanz, Nucl. Phys. A {\bf 688},  430c
  (2001).

\bibitem{Amari2002}
S. Amari, New Astron. Rev. {\bf 46},  519  (2002).

\bibitem{Jose2004}
J. Jos{\'{e}} {\it et~al.}, Astrophys. J. {\bf 612},  414  (2004).

\bibitem{Starrfield2016}
S. Starrfield, C. Iliadis, and W.~R. Hix, {Publications of the
Astronomical Society of the Pacific}, {\bf 128}, 051001 (2016).

\bibitem{Shafter2017}
A.~W. Shafter, Astrophys. J. {\bf 834},  196  (2017).

\bibitem{Iliadis2002}
C. Iliadis {\it et~al.}, Astrophys. J. Suppl. Ser. {\bf 142},  105  (2002).

\bibitem{Runkle2001}
R.~C. Runkle, A.~E. Champagne, Engel, and J, The Astrophysical Journal, {\bf 556}, 970-978, (2001).

\bibitem{Peplowski2009}
P.~N. Peplowski {\it et~al.}, Phys. Rev. C {\bf 79},  032801(R)  (2009).

\bibitem{Burlein1984}
M. Burlein, K.~S. Dhuga, and H.~T. Fortune, Phys. Rev. C {\bf 29}, 2013
  (1984).

\bibitem{Arciszewski1984a}
H.~F. Arciszewski {\it et~al.}, Nucl. Physics, Sect. A {\bf 430},  234
  (1984).

\bibitem{Richter2011}
W.~A. Richter, B.~A. Brown, A. Signoracci, and M. Wiescher, Phys. Rev. C {\bf
  83},  065803  (2011).

\bibitem{Yasue1990a}
M. Yasue {\it et~al.}, Phys. Rev. C {\bf 42},  1279 (1990).

\bibitem{Parikh2013}
A. Parikh and J. Jos{\'{e}}, Phys. Rev. C {\bf 88},  048801  (2013).

\bibitem{Marshall2019}
C. Marshall {\it et~al.}, IEEE Trans. Instrum. Meas. {\bf 68},  533  (2019).

\bibitem{Basunia2016}
M.~S. Basunia {\it et~al.}, Nucl. Data Sheets {\bf 134},  1  (2016).

\bibitem{Koning2003}
A.~J. Koning and J.~P. Delaroche, Nucl. Phys. A {\bf 713},  231  (2003).

\bibitem{Han2006}
Y. Han, Y. Shi, and Q. Shen, Phys. Rev. C {\bf 74}, 044615,  (2006).

\bibitem{Soukhovitski2005}
E.~S. Soukhovitski, S. Chiba, and J.~Y. Lee, AIP Conf. Proc. {\bf 769},  1100
  (2005).

\bibitem{Thompson1988}
I. Thompson, Comput. Phys. Reports {\bf 7},  167  (1988).

\bibitem{Austern1987}
N. Austern {\it et~al.}, Phys. Rep. {\bf 154},  125  (1987).

\bibitem{Avrigeanu2010}
M. Avrigeanu and V. Avrigeanu, J. Phys. Conf. Ser. {\bf 205},  012014 (2010).

\bibitem{Kamimura1986}
M. Kamimura {\it et~al.}, Prog. Theor. Phys. Suppl. {\bf 89},  1  (1986).

\bibitem{Nguyen2010}
N.~B. Nguyen, F.~M. Nunes, and R.~C. Johnson, Phys. Rev. C {\bf 82},  014611,  (2010).

\bibitem{An2006}
H. An and C. Cai, Phys. Rev. C {\bf 73},  054605 (2006).

\bibitem{Li2008}
X. Li and C. Cai, Nucl. Phys. A {\bf 801},  43  (2008).

\bibitem{Koning2007}
A.~J. Koning, S. Hilaire, and M.~C. Duijvestijn, Int. Conf. Nucl. Data Sci.
  Technol.  211  (2008).

\bibitem{Hodgson1971}
P.~E. Hodgson,  in {\em {Nuclear Reactions and Nuclear Structure}}, edited by
  W. Marshall and D. Wilkinson (Oxford University Press, London, 1971), p.\
  302.

\bibitem{Gallmann1966}
A. Gallman {\it et~al.}, Nucl. Phys. {\bf 88},  654  (1966).

\bibitem{Schmick1974}
T.~A. Schmick, K.~W. Kemper, P.~K. Bindal, and R.~D. Koshel, Phys. Rev. C {\bf
  10},  556  (1974).

\bibitem{Meurders1975}
F. Meurders and G. {De Korte}, Nucl. Phys. A {\bf 249},  205  (1975).

\bibitem{Huang2017}
W.~J. Huang {\it et~al.}, Chinese Phys. C {\bf 41},  030002  (2017).

\bibitem{DeSereville2010}
N. {De S{\'{e}}r{\'{e}}ville} {\it et~al.},   Proc. Sci., NIC XI, 212, (2010).

\bibitem{Komatsubara2014}
T. Komatsubara {\it et~al.}, Eur. Phys. J. A {\bf 50},  1  (2014).

\bibitem{Doherty2015a}
D.~T. Doherty {\it et~al.}, Phys. Rev. C {\bf 92}, 035808  (2015).

\bibitem{Parpottas2004}
Y. Parpottas {\it et~al.}, Phys. Rev. C {\bf 70},  065805  (2004).

\bibitem{Matic2010}
A. Matic {\it et~al.}, Phys. Rev. C {\bf 82},  025807  (2010).

\bibitem{Iliadis1996}
C. Iliadis {\it et~al.}, Phys. Rev. C, {\bf 53},  1 (1996).

\end{thebibliography}
\end{document}